\documentstyle[aps,preprint,tighten,epsfig,floats]{revtex}

\begin{document}

\draft
\preprint{LA-UR-98-3087}
\title{Continuous Probability Distributions from Finite Data}
\author{David M. Schmidt}
\address{Biophysics Group, Los Alamos National Laboratory, 
          Los Alamos, New Mexico 87545}
\date{August 5, 1998}
\maketitle
\begin{abstract}
  Recent approaches to the problem of inferring a continuous
  probability distribution from a finite set of data have used a
  scalar field theory for the form of the prior probability
  distribution.  This letter presents a more general form for the
  prior distribution that has a geometrical interpretation which is
  useful for tailoring prior distributions to the needs of each
  application.
  Examples are presented that demonstrate some of the
  capabilities of this approach, including the applicability of this
  approach to problems of more than one dimension.          
\end{abstract}


Inferring the continuous probability distribution, or target
distribution, from which a finite number of data samples were drawn is
an example of an ill-posed inverse problem: there are many different 
distributions that could have produced the given finite data.  Often
one has prior information, separate from the data itself, that can
reduce the range of possible target distributions.  More
generally, one can assign a prior probability to each target 
distribution based on the prior information.  Combining this prior
probability distribution with the likelihood of the data given any
particular target distribution, using Bayes' rule of probability,
produces a posterior probability over the space of target
distributions.  This posterior distribution encapsulates all the
information available, both from the data and from the prior
information, and can be used to make probabilistic inferences.

Let $P[Q|x_1,\ldots,x_N]$ denote the posterior probability that the
target distribution $Q(x)$
describes the data $x_1,\ldots,x_N$.   By Bayes' rule,
\begin{eqnarray}
\label{eq:Bayes}
P[Q|x_1,\ldots,x_N] &=& {P[x_1,\ldots,x_N|Q] P[Q]\over
P[x_1,\ldots,x_N]} \\
&=& {Q(x_1)\cdots Q(x_N) P[Q] \over \int {\cal D}Q \, Q(x_1)\cdots
Q(x_N) P[Q]},
\end{eqnarray}
where $P[Q]$ is the prior probability of the target distribution
$Q$.

The form for $P[Q]$ should incorporate the available prior
information.  For example, by setting $Q(x) = \psi^2(x)$
\cite{Good_Gaskins_71}, where $\psi$ may take any value in
$(-\infty,\infty)$, we may insure that $Q$ is non-negative.  $\psi$ is
referred to as the {\em amplitude\/} by analogy with quantum mechanics
\cite{holy}.  A particular form for $P[Q]$, or rather $P[\psi]$, that
has been presented in order to, the authors say, incorporate a bias
that $Q$ be ``smooth'' is \cite{holy,BCS,Periwal}
\begin{equation}
\label{eq:Ppsi}
P[\psi] = {1\over Z} \exp\left[- \int dx\, {\ell^2\over2}
(\partial_x\psi)^2 \right] \delta\left(1-\int dx\, \psi^2\right),
\end{equation}
where $Z$ is the normalization factor and $\ell$ is a constant
which controls the penalty applied to gradients.  The delta function
enforces normalization of the distribution $Q$.

Because this particular prior distribution is not very effective at
generating smooth distributions (as will be shown) and because the
prior information available for each problem will vary, it is useful
to consider a more general form for the prior distribution.  A more
general approach is to define the prior distribution as
\begin{equation}
\label{eq:PpsiV}
P[\psi] = {1\over Z} \exp\left[- {1\over2} \langle \psi | V^{-1} |
  \psi \rangle\right] \delta\left(1-\langle
  \psi | \psi \rangle\right),
\end{equation}
where $V$ is a positive, symmetric (Hermitian) operator within
whatever Hilbert space is
chosen for $\psi$.  This distribution is a generalization of a
multi-dimensional Gaussian distribution with $V$ acting as the
covariance operator.  Continuing this analogy, we write
\begin{equation}
\label{eq:covvar}
V(\mathbf{x},\mathbf{y}) =
\sigma(\mathbf{x})\sigma(\mathbf{y})\rho(\mathbf{x},\mathbf{y})
\end{equation}
where $\sigma^2(\mathbf{x})$ is the variance at $\mathbf{x}$ and
$\rho$ is the correlation function.  Information about smoothness is
encoded in the correlation function.  For example, if the distribution
from which the $\{\mathbf{x}_i\}$ were drawn is expected to be smooth
over distances smaller than a certain spatial scale then the
correlation function should be near unity over distances smaller than
this scale. The prior distribution used in \cite{holy,BCS} is
equivalent to the one presented here in one dimension with $V^{-1} = -\ell^2
\partial_x^2$, assuming $\psi$ goes to zero at $\pm \infty$.

It is useful to consider this prior probability distribution in
geometrical terms.  The eigenfunctions of the the operator $V$ form a
basis for the space of $\psi$.  The normalization constraint restricts
$\psi$ to lie on a hyper-spherical surface of radius one.  Those
eigenfunctions with larger eigenvalues are more likely, \emph{a
  priori}. If $V$ has any eigenvalues that are zero then the
corresponding eigenfunctions form a basis for a subspace that is
orthogonal to $\psi$; that is the prior distribution prevents $\psi$
from having any components along
these eigenfunctions.

With this form for the prior distribution the probability
$P[Q|x_1,\ldots,x_N]$ of a distribution $Q$ given the data is
\begin{eqnarray}
\lefteqn{P[\psi|x_1,\ldots,x_N] \propto \psi^2(x_1) \cdots
\psi^2(x_N)}\qquad&&\nonumber\\ & & \times \exp\left[-{1\over2}\langle
\psi | V^{-1} | \psi \rangle\right] \delta\left(1-\langle
\psi | \psi \rangle\right) \\ &=& e^{-S[\psi]} \delta\left(1-\langle
\psi |  \psi \rangle\right),
\label{eq:Pofpsi}
\end{eqnarray}
where
the effective action $S$ is
\begin{equation}
S[\psi] = {1\over2}\langle
\psi | V^{-1} | \psi \rangle - 2\sum_i \ln\left(\langle x_i | \psi \rangle\right).
\end{equation}

The most likely distribution given the data is that function
$\psi_{\rm cl}$ which minimizes the effective action subject to the
normalization constraint.  To enforce this constraint a Lagrange
multiplier term $\lambda(1-\langle \psi | \psi \rangle)/2$ is
subtracted from the action.  Variational methods then lead to the
following equations for $\psi_{\rm cl}$ and $\lambda$:
\begin{mathletters}
\label{eq:qcl}
\begin{equation}
\label{eq:qcla}
|\psi_{\rm cl}\rangle = 2\sum_i
\frac{(V^{-1}+\lambda I)^{-1}|x_i\rangle}{\langle
  x_i | \psi_{\rm cl} \rangle}
\end{equation}
\begin{equation}
\label{eq:qclnorm}
\langle \psi_{\rm cl} | \psi_{\rm cl}\rangle = 1.
\end{equation}
\end{mathletters}
The solution to these equations may be written
\begin{equation}
\label{eq:qcldef}
|\psi_{\rm cl}\rangle = \sum_i a_i U(\lambda)|x_i \rangle,
\end{equation}
where $U(\lambda) = (V^{-1}+\lambda I)^{-1}$.  
Eqs.~(\ref{eq:qcl}) imply
\begin{mathletters}
\label{eq:alam}
\begin{equation}
\label{eq:a1}
a_i\sum_j a_j \langle x_i | U(\lambda) | x_j\rangle =
2, \qquad i = 1,\ldots,N
\end{equation}
\begin{equation}
\label{eq:a2}
\sum_{i,j} a_ia_j \langle x_i | U^2(\lambda) | x_j\rangle = 1.
\end{equation}
\end{mathletters}
These $N+1$ non-linear equations determine $\lambda$ and the $a_i$ and
may be solved using Newton's method \cite{holy}.

\begin{figure}
\centerline{\epsfig{file=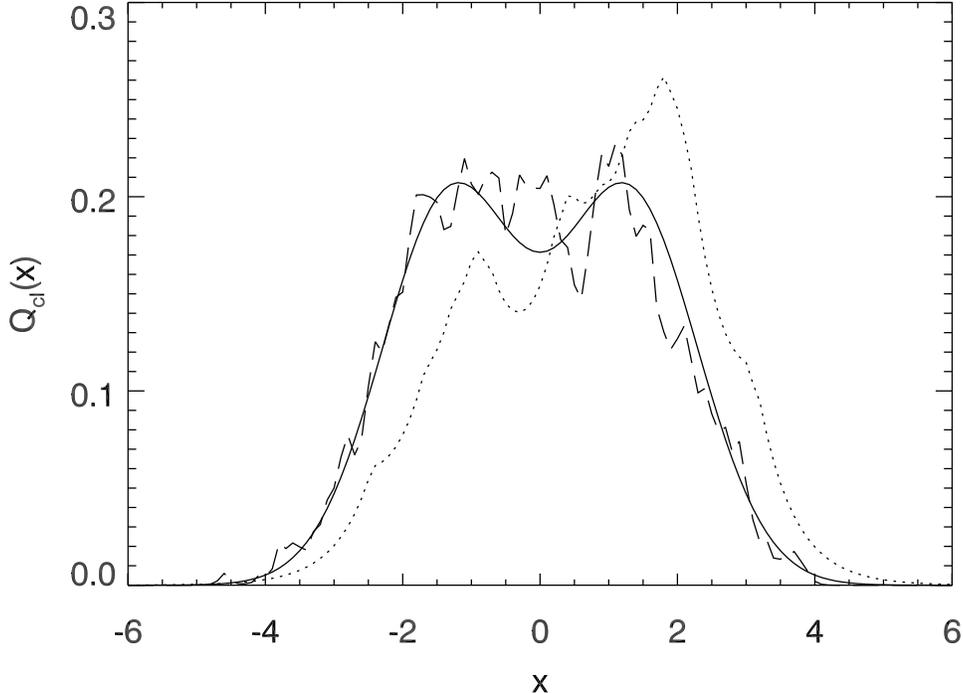,width=5in,angle=0}}
\caption{
\label{fig:helm}
The most likely distributions from an inverse Laplacian prior
distribution with $\ell = 6$ and from $N = 20$ (dashed line) and $N =
1000$ (dotted line) data drawn randomly from a target distribution
consisting of the sum of two Normal distributions (solid curve).}  
\end{figure}

The covariance operator $V$ in the prior distribution should be chosen
for each different probability distribution that one is estimating.  A
few examples with three different forms for $V$ are described below in
order to illustrate the effects that different choices of $V$ can
have.  First consider the case used in \cite{holy,BCS} in which the
prior covariance operator is an inverse Laplacian in one dimension, $V^{-1} = -\ell^2
\partial_x^2$.  In this case
\begin{equation}
\label{eq:uc1}
U(\lambda) = \left( -\ell^2 \partial_x^2 + \lambda I\right)^{-1}
\end{equation}
which is the Green's function of the modified Helmholtz equation.  The
solutions of this equation are well known, even for dimensions larger
than one \cite{Arfken}.  In particular, in one dimension the
most likely solution $\psi_{\rm cl}(x)$ is, from Eq.~\ref{eq:qcldef}
\begin{equation}
\label{eq:psicl1}
\psi_{\rm cl}(x) = \sum_i a_i \frac{1}{2k\ell^2}\exp\left(-k|x-x_i|\right)
\end{equation}
where $k = \sqrt{\lambda}/\ell$. 
Examples of the most likely probability distributions for this case
with $\ell = 6$ are shown in Fig.~\ref{fig:helm}.  For these examples
the data were drawn from a target distribution consisting of the sum
of two Normal distributions, shown as the solid curve in the figure.
The most
likely distributions are not
very smooth, as would be expected from the functional form of
Eq.~\ref{eq:psicl1}.

For the second example consider the case in which the prior covariance
operator has a correlation function which is a Gaussian,
\begin{equation}
\label{eq:vn2}
V(\mathbf{x},\mathbf{y};\mathnormal{r})=\sigma^2\exp\left[\frac{-(\mathbf{x}-\mathbf{y})^2}{2 r^2}\right].
\end{equation}
Here $\sigma^2$ is the prior variance for the magnitude of the target
probability distribution and $r$ is a correlation scale below
which the target probability distribution is believed to be smooth.
In this case it is useful to expand $U$ in an operator product
expansion in $V$,
\begin{equation}
\label{eq:uope}
U(\lambda) = V\left( 1 - \lambda\, V + \lambda^2\, V\cdot V - \lambda^3\,
  V\cdot V\cdot V + \cdots \right).
\end{equation}
Because $V\cdot V \propto
V(\mathbf{x},\mathbf{y};\mathnormal{\sqrt{2}r})$ for this particular
$V$, Eq.~\ref{eq:uope} generates a multi-resolution expansion,
analogous to a wavelet expansion, for $U$ and therefore also for
$\psi_{\rm cl}$ consisting of Gaussians of ever increasing 
width, increasing each step by a factor of $\sqrt{2}$ with the finest
scale being represented by the original
$V(\mathbf{x},\mathbf{y};\mathnormal{r})$.  This functional form for $V$
therefore generates a most likely probability distribution that has
finite derivatives to all orders and is generally more smooth than
that from the first example.

\begin{figure}
\centerline{\epsfig{file=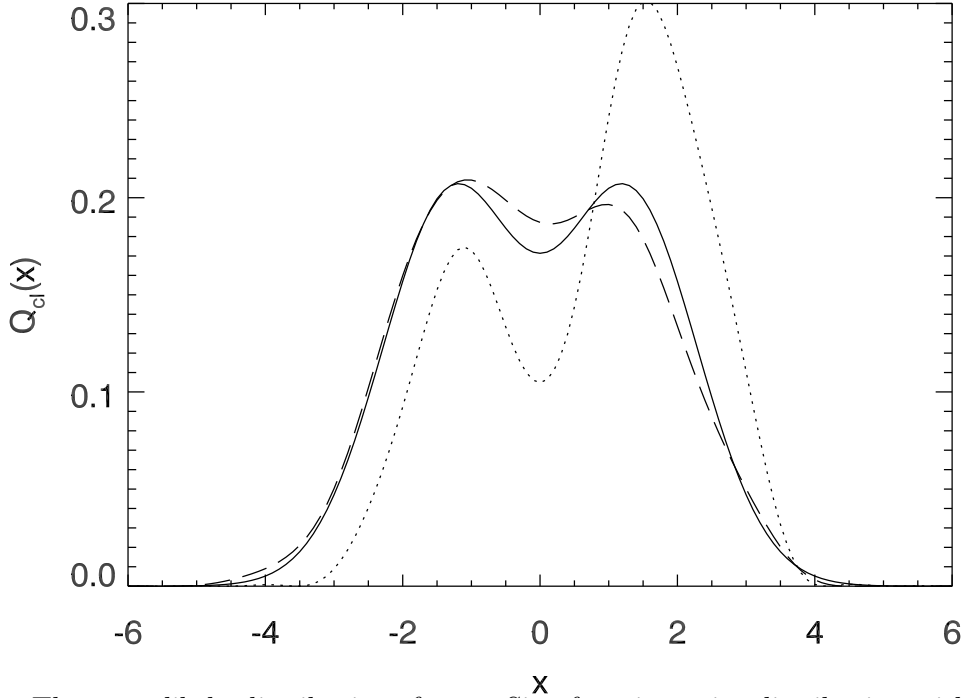,width=5in,angle=0}}
\caption{
\label{fig:result3}
The most likely distributions from a Sinc function prior distribution
with $k_0 = 3.33$ and from the same $N = 20$ (dashed line) and $N =
1000$ (dotted line) data used for the examples in Fig.~\ref{fig:helm},
which were drawn randomly from a target distribution consisting of the
sum of two Normal distributions (solid curve).}
\end{figure}

For the final example, consider the case in which the prior covariance
operator is a projection operator that projects onto the subspace
formed by functions having only Fourier wavenumbers smaller than a
particular wavenumber $k_0$.  In one dimension this covariance
operator is the Sinc function,
\begin{equation}
\label{eq:v3p}
V(x,y;k_0) = \frac{\sin\left[k_0(x-y)\right]}{\pi(x-y)}.
\end{equation}
Because this is a projection operator, $V\cdot V = V$ and from
Eq.~\ref{eq:uope} $U$ for this case is simply $U(\lambda)=V/(1 +
\lambda)$. The most likely amplitude therefore consists of sums of
Sinc functions centered at each data point.  Examples of the most
likely probability distribution using this prior distribution with
$k_0=3.33$ are shown in Fig.~\ref{fig:result3}.  The same data used
for the examples in Fig.~\ref{fig:helm} were used here.  Even with
only 20 data points the most likely solution indicates a doubly peaked
distribution.  Both of the examples here are more smooth than those
generated by the prior distribution discussed above in the first case
and shown in Fig.~\ref{fig:helm}.

\begin{figure}
\centerline{\epsfig{file=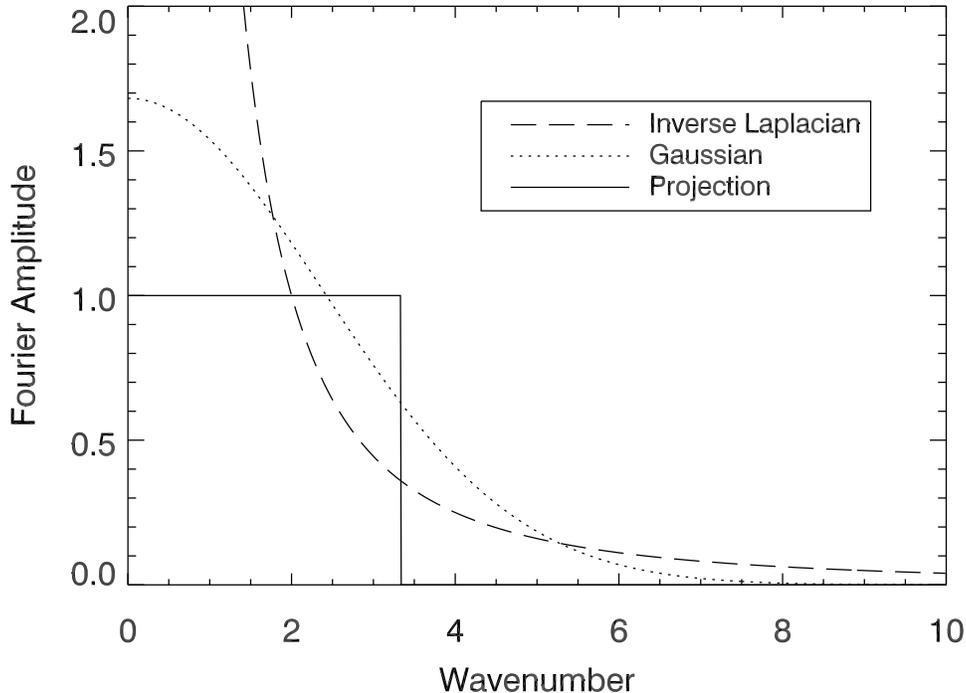,width=5in,angle=0}}
\caption{
\label{fig:forspe}
  The Fourier spectra of the three types of covariance operators shown
  in the legend.  Free parameters in each case have been set to
  correspond roughly to a cutoff at wavenumber $k_0=3.33$.  }
\end{figure}

It is useful to examine the Fourier spectrum of the prior covariance
operator in order to understand some of the properties of the resulting
most likely distribution.  The Fourier spectra of the three covariance
operators considered in the above examples are shown in
Fig.~\ref{fig:forspe}. Because of the form of the prior distribution
(Eq.~\ref{eq:PpsiV}) those wavenumbers with larger Fourier
amplitudes are more likely, \emph{a priori}.  However, in order to
maximize the likelihood of the given data, the most likely amplitude
will tend to consist of the largest possible wavenumber 
components.  Because the Sinc function covariance operator has the
sharpest high wavenumber cutoff it will tend to generate the smoothest
most likely distribution.  Conversely, the inverse Laplacian covariance
operator will tend to produce the least smooth most likely
distribution.  If the Sinc function covariance operator is used,
however, the cutoff $k_0$ should be chosen with great care because
this prior forbids any solutions containing wavenumbers higher than
the cutoff.  Thus if the chosen cutoff wavenumber is lower than
the maximum wavenumber component of the target distribution then
the most likely distribution will not converge to the target
distribution as the number of data points increases.

\acknowledgements
Supported by Los Alamos National Laboratory and by NIDA/NIMH Grant
DA/MH09972, J.S. George, Principal Investigator.


\end{document}